\documentclass{natureprintstyle}
\usepackage{aastex_macros}
\usepackage{hyperref}
\usepackage{graphicx}
\usepackage{amsmath}
\usepackage{amssymb}
\usepackage{lineno}
\usepackage{color}

\newcommand{\Mjup}[1]{${#1}~M_{\rm Jup}$}

\newcommand{\vel}[1]{${#1}~{\rm m\,s^{-1}}$}

\renewcommand{\apj}{Astrophys. J.}

\renewcommand{\apjs}{Astrophys. J. Supp.}

\renewcommand{\aap}{Astron. Astrophys.}
\renewcommand{\aj}{Astron. J.}

\renewcommand{\aaps}{A\&AS}

\title{Meridional flows in the disk around a young star}

\author{Richard Teague$^{*, 1}$,
	    Jaehan Bae$^{2}$,
        Edwin A. Bergin$^{1}$}

\begin{document}

\maketitle

\let\thefootnote\relax\footnote{
\begin{affiliations}

\item Department of Astronomy, University of Michigan, 311 West Hall, 1085 S. University Ave, Ann Arbor, MI 48109, USA

\item Department of Terrestrial Magnetism, Carnegie Institution for Science, 5241 Broad Branch Road, NW, Washington, DC 20015, USA

\end{affiliations}
}


\vspace{-3.3mm}
\begin{abstract}
Protoplanetary disks are known to posses a stunning variety of substructure in the distribution of their mm~sized grains, predominantly seen as rings and gaps\cite{Andrews_ea_2018}, which are frequently interpreted as due to the shepherding of large grains by either hidden, still-forming planets within the disk\cite{Zhang_ea_2018} or (magneto-)hydrodynamic instabilities \cite{Flock_ea_2015}. The velocity structure of the gas offers a unique probe of both the underlying mechanisms driving the evolution of the disk, the presence of embedded planets and characterising the transportation of material within the disk, such as following planet-building material from volatile-rich regions to the chemically-inert midplane, or detailing the required removal of angular momentum. Here we present the radial profiles of the three velocity components of gas in upper disk layers in the disk of HD~163296 as traced by $^{12}$CO molecular emission. These velocities reveal significant flows from the disk surface towards the midplane of disk at the radial locations of gaps argued to be opened by embedded planets\cite{Isella_ea_2016, Isella_ea_2018, Teague_ea_2018a, Pinte_ea_2018b}, bearing striking resemblance to meridional flows, long predicted to occur during the early stages of planet formation\cite{Szulagyi_ea_2014, Morbidelli_ea_2014, Fung_Chiang_2016, Dong_ea_2019}. In addition, a persistent radial outflow is seen at the outer edge of the disk, potentially the base of a wind associated with previously detected extended emission\cite{Klaassen_ea_2013}.
\end{abstract}

We use observations of $^{12}$CO $J = 2-1$ emission from HD~163296 to measure the 3D velocities structure of the gas. These data were originally presented as part of the Disk Substructures at High Angular Resolution Project (DSHARP) Atacama Large Millimeter/submillimeter Array (ALMA) large program\cite{Andrews_ea_2018, Isella_ea_2018}, which combined previously analysed lower spatial resolution data\cite{Flaherty_ea_2015, Isella_ea_2016}. The disk around HD~163296 is known to host multiple rings of large mm~sized grains trapped within regions of gas pressure maxima\cite{Teague_ea_2018a}, with a depletion of the gas within the dust gaps\cite{Isella_ea_2016}, highly suggestive of a planetary origin. In the outer disk, at a radius of $\approx 260$~au (where an astronomical unit, au, is the distance between the Earth and the Sun), a local disturbance in the velocity field is likely driven by a massive, \Mjup{2} planet (where \Mjup{} is the mass of Jupiter), deeply embedded within the disk\cite{Pinte_ea_2018b}.

\begin{figure*}
    \centering
    \includegraphics[width=\textwidth]{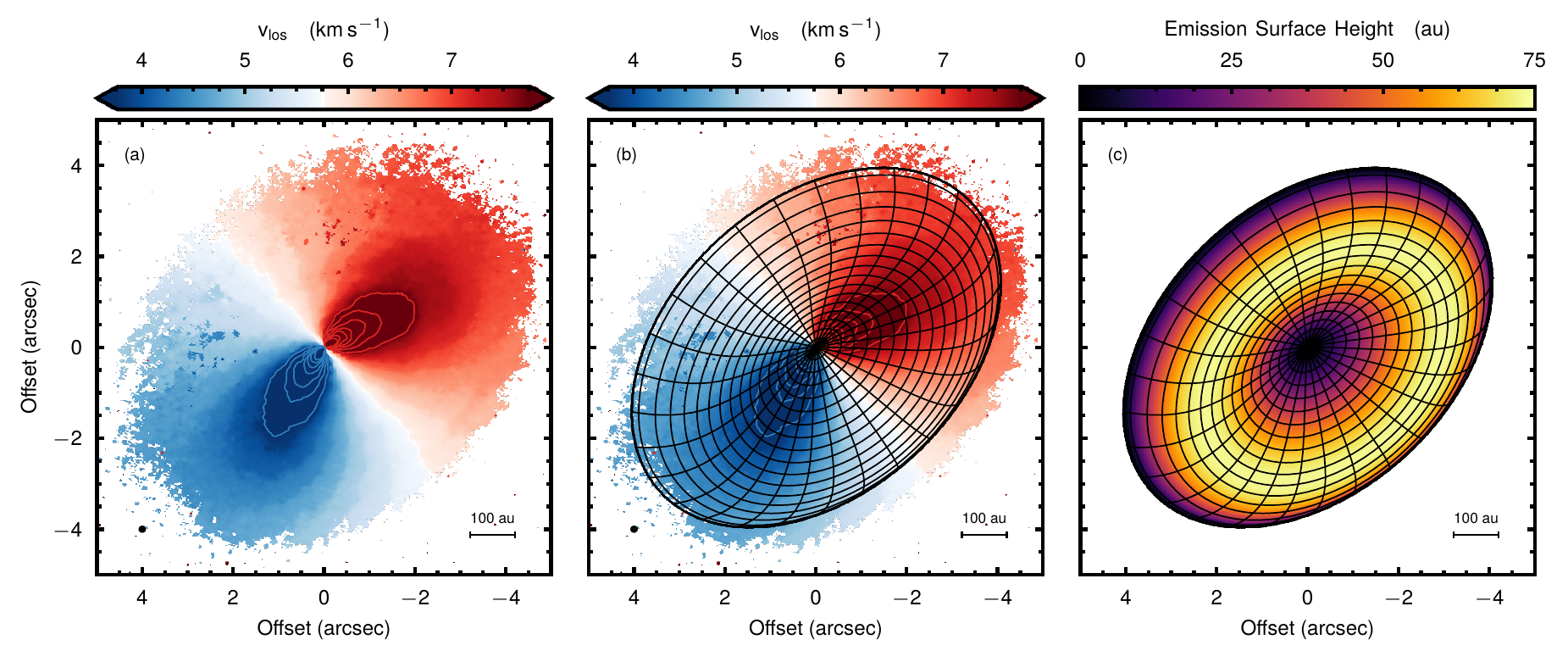}
    \caption{\textbf{Rotation maps and the inferred 3D geometry of the disk.} Panel (a) shows the rotation maps of HD~163296 made with \texttt{bettermoments}\cite{Teague_Foreman-Mackey_2018}. The filled contours have been clipped to highlight the structure in the disk close to the semi-minor axis. The lined contours are in steps of $0.5~{\rm km\,s^{-1}}$. Panel (b) shows an overlay of the inferred emission surface used to deproject the data, with panel (c) showing the inferred height above the midplane. The synthesised beam is shown in the bottom left of the two left panels as an ellipse.}
    \label{fig:12COv0}
\end{figure*}

To improve the signal to noise and achieve high velocity precision we first radially bin the spectral data\cite{Teague_ea_2018a}. As $^{12}$CO $J=2-1$ is optically thick, the $\tau \approx 1$ surface traces a region typically 2 -- 4 pressure scale heights above the disk midplane, resulting in asymmetries in the observed emission profiles which must be taken into account when radially binning the data\cite{Rosenfeld_ea_2013, Pinte_ea_2018a, Isella_ea_2018}. This is most clearly seen in a map of the line center, shown in the left panel of Fig.~\ref{fig:12COv0}, which deviates significantly from the symmetric dipole pattern found in geometrically thin disks. As Keplerian rotation dominates the velocity structure, a parametric emission surface can be inferred which correctly accounts for this projection effect, with the best-fit surface shown in Fig.~\ref{fig:12COv0}. This approach finds an emission surface consistent with another technique which yields a non-parametric surface\cite{Pinte_ea_2018a, Teague_ea_2018a}; the latter, however, is highly sensitive to noise in the data and so can only poorly constrain the emission surface in the outer disk.

Azimuthally averaged gas velocities are measured by first splitting the disk into annuli of constant radius using the inferred emission surface. Then, for each radius, we use the azimuthal dependence of the Doppler shift of the line centres due to the projection of the gas velocities and infer the azimuthally averaged rotational and radial velocities, $v_{\phi}$ and $v_{R}$\cite{Teague_ea_2018a, Teague_ea_2018b}. As the $v_{\phi}$ and $v_{R}$ components are orthogonal to one another, their projected components have a different azimuthal dependence and can therefore be readily disentangled. Deviations in the line centre at a specific radius relative to the systemic velocity, $v_{\rm LSR}$, is interpreted as a vertical component, $v_{Z}$. This approach extends the method in Teague et al. (2018a,b)\cite{Teague_ea_2018a, Teague_ea_2018b} which considered only $v_{\phi}$ components. This improved method recovers $v_{\phi,\, {\rm proj}}$ and $v_{R,\, {\rm proj}}$ absolutely, while $v_{Z,\, {\rm proj}}$ and the correction into disk-frame velocities, $(v_{\phi}, \, v_{R}, \, v_{Z})$, requires a precise measurement of $v_{\rm LSR}$ and the disk inclination, $i$, respectively. The former is well constrained from measurement of the emission surface, $v_{\rm LSR} = 5763 \pm 1~{\rm m\,s^{-1}}$, while the disk inclination is taken from fits to the rings in continuum\cite{Isella_ea_2018} which find a disk average $i = 46.7\degr \pm 0.1\degr$. The resulting statistical uncertainties on $v_{\phi}$ and $v_{R}$ are $\sim$~\vel{10} while for $v_{Z}$ they are $\sim$~\vel{20} due to the additional step of measuring the line center. We estimate the systematic uncertainties associated with the choice of emission surface are equivalent to between 2 and 3 times the statistical uncertainties. However, as discussed in the Methods section, a different emission surface does not significantly change the observed flow structure.

\begin{figure*}
    \centering
    \includegraphics[width=\textwidth]{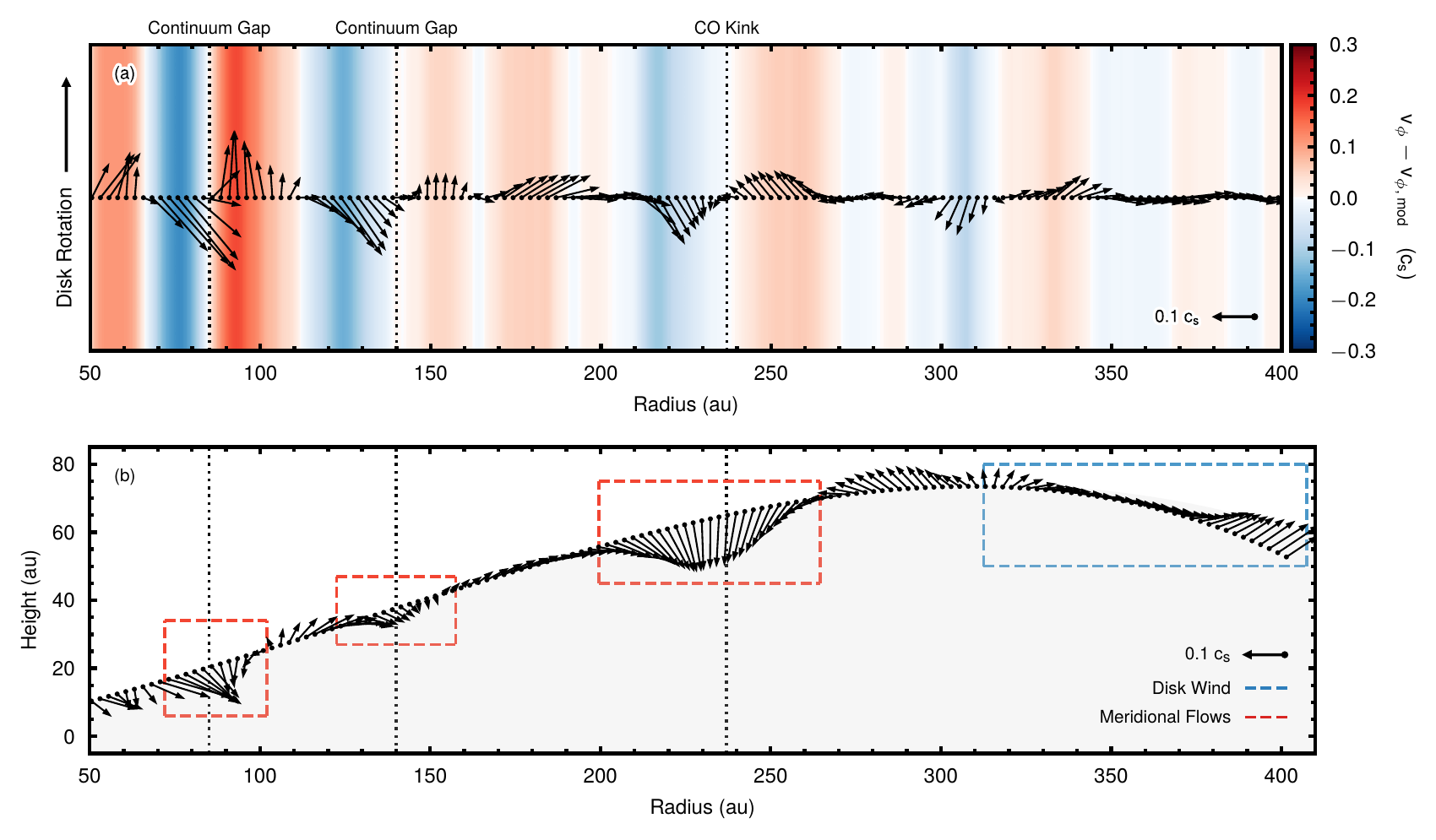}
    \caption{\textbf{Measured velocity structure for the gas in the disk around HD~163296.} Inferred 3D velocity structure in the $(R,\,\phi)$ plane, (a), with the disk rotation along the vertical axis, and the $(R,\,Z)$ plane, (b). All velocities are deprojected assuming a source inclination of $46.7\degr$ and converted to a fraction of the local sound speed, $c_s$. In both panels a vector in the bottom right shows $0.1~c_s$. In (a), vectors in the positive and negative $y$ direction faster or slower-rotating material, respectively, while the $x$ direction shows the radial flows. The colour background shows the magnitude of the rotation vector. In panel (b), the three locations of meridional circulation are shown in red dashed boxes and the outflow in a blue dashed box. The dotted lines mark the locations of the gaps in the continuum emission\cite{Huang_ea_2018b, Isella_ea_2018}, and the local velocity disturbances traced in $^{12}$CO emission\cite{Pinte_ea_2018b}.}
    \label{fig:12COvectors}
\end{figure*}

To relate the gas velocities to the local sound speed, we use the brightness temperature of the optically thick line emission as a measure the local gas temperature\cite{Isella_ea_2018}, finding values spanning between $\approx 90$~K in the inner 30~au and dropping to 30~K at 400~au. This results in local sound speeds ranging between \vel{600} and \vel{300}. As the pressure supported Keplerian rotation dominates $v_{\phi}$, we subtract a baseline model, $v_{\phi,\, {\rm mod}}$, in order to highlight local deviations in $v_{\phi}$. Fig.~\ref{fig:12COvectors} shows the resulting velocity structure normalised to the local sound speed in both the $(R,\,\phi)$ plane and the $(R,\,Z)$ plane. We find striking evidence of a highly dynamic disk. In particular, we find three regions of a collapsing flow (Fig.~\ref{fig:12COvectors}b) with gas rotating at slower and faster velocities either side of the maximum $|v_Z|$ (Fig.~\ref{fig:12COvectors}a), coincident in radius with three previously claimed planets and gaps in the dust continuum at 87, 140 and 237~au\cite{Isella_ea_2016, Isella_ea_2018, Pinte_ea_2018b, Teague_ea_2018a}. In addition we observe a persistent radially outwards flow beyond $\sim$300~au.

\begin{figure*}
    \centering
    \includegraphics[width=0.9\textwidth]{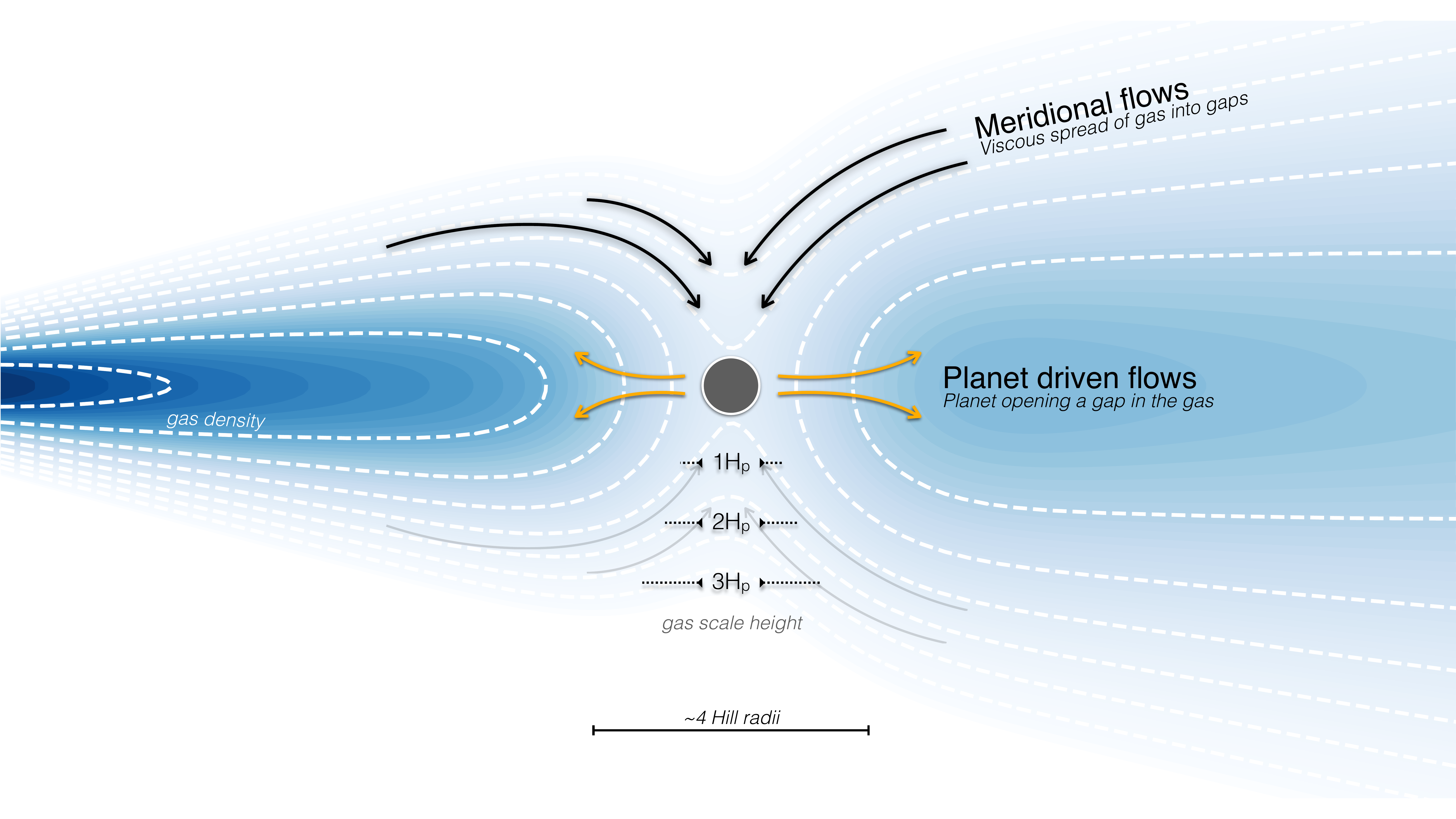}
    \caption{\textbf{Schematic of the meridional flow.} Cartoon of the dominant flows around an embedded planet after Fung \& Chiang (2016). The planet will open a gap in the gas density through Lindblad torques driving the radial flows shown in orange close to the midplane. The disk will viscously spread towards the gap center, then fall into the gap via meridional flows, shown with black arrows at higher altitudes. For wide gaps, this will result in circulation around the gap edges.}
    \label{fig:cartoon}
\end{figure*}

The collapsing regions are bounded by annuli of negative and positive residuals of $v_{\phi} - v_{\phi,\, {\rm mod}}$, indicative of a pressure minimum and demonstrating the presence of a significant gap in the gas surface density\cite{Teague_ea_2018a, Teague_ea_2018b}. Gas is seen to flow towards the gap centres before falling into the region of low gas pressure. A promising candidate for driving such flow structures are meridional flows, which have long been predicted from 3D hydrodynamic simulations at the radii of embedded planets with their origin well understood\cite{Kley_ea_2001, Szulagyi_ea_2014, Morbidelli_ea_2014, Fung_Chiang_2016, Dong_ea_2019}. As seen schematically in Fig.~\ref{fig:cartoon}, embedded planets will open gaps in the gas by driving material away from the planet at the midplane via Lindblad torques\cite{Lin_Papaloizou_1986}. Locally, the gas pressure drops causing a decrease in the local gas scale height in order for the disk to maintain vertical hydrostatic equilibrium. In turn, this creates a region of low pressure at the disk surface, at heights as high as 3 -- 4 pressure scale heights above the midplane. Gas from surrounding regions of higher pressure will therefore flow towards the region of lower pressure at the gap center, before falling towards the midplane. It is this gas near the disk surface which is entrained into the flow towards the gap center and then downwards to the midplane that we trace with optically thick $^{12}$CO observations.

\begin{figure*}
    \centering
    \includegraphics[width=\textwidth]{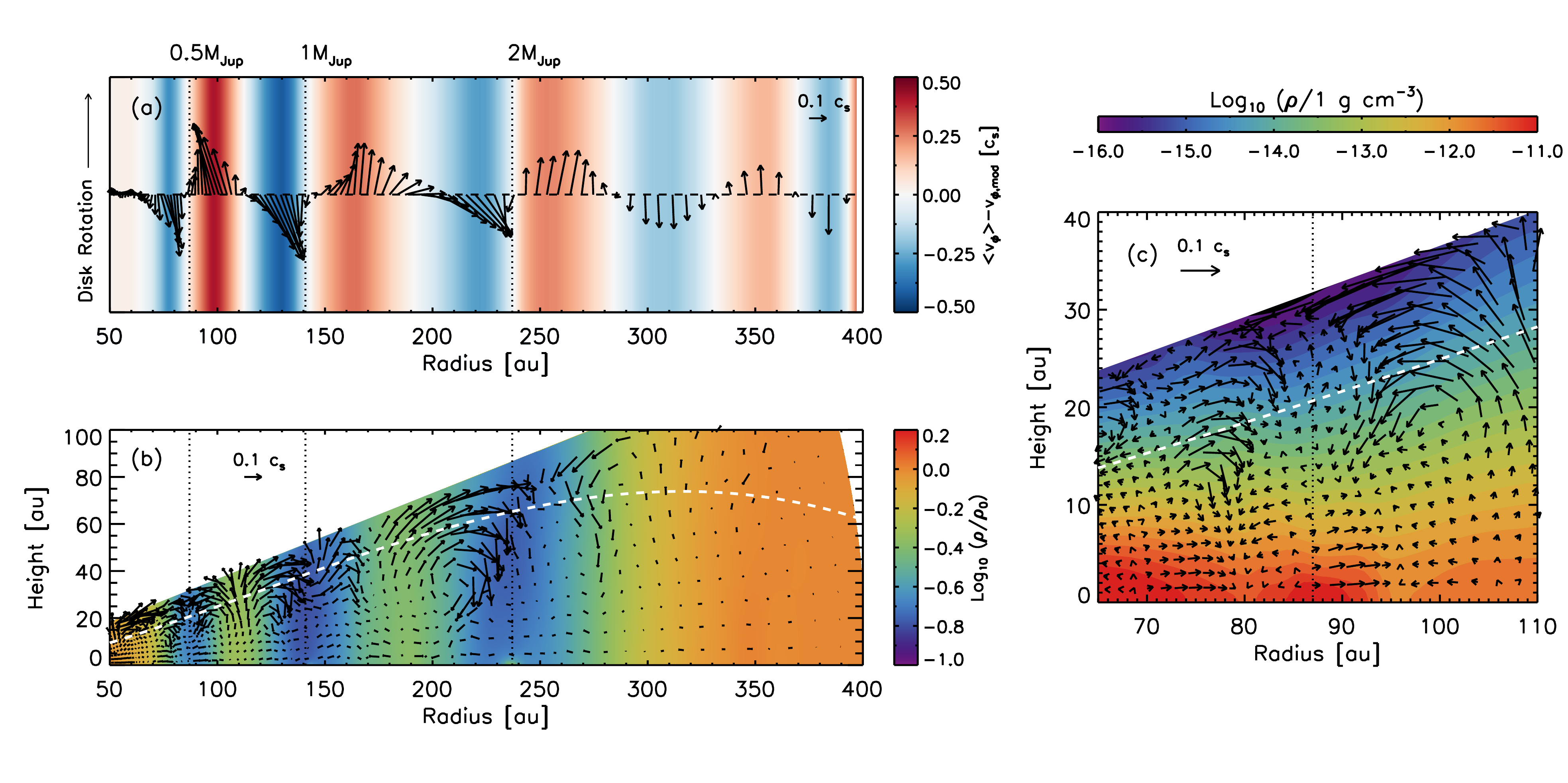}
    \caption{\textbf{Hydrodynamical simulations of meridional flows.} Azimuthally averaged 3D velocity structure (a) in the $(R, \phi)$ plane at the $^{12}$CO emission surface and (b) in the $(R, Z)$ plane, similar to Figure \ref{fig:12COvectors} but from the hydrodynamic simulation. The colour contours in panel (b) show the disk gas density, normalised by its initial value in each grid cell. The white dashed curve in panel (b) presents the inferred $^{12}$CO emission surface as in Equation \ref{eqn:emission_surface}. In panel (c), we present a zoom-in view of the velocity field at the vicinity of the planet at 87~au. The disk gas density along the R-Z slice containing the planet is shown in the background in a logarithmic scale. All the contours and vectors show snapshot values taken at the end of the simulation (1.44~Myr). For visualisation purpose only, we present vectors every two radial and meridional grid cells in panels (a) and (c), and every three radial and meridional grid cells in panel (b).}
    \label{fig:hydro}
\end{figure*}

We run a 3D hydrodynamical simulation of embedded planets in order to provide a qualitative comparison to the observations and demonstrate that $^{12}$CO emission would trace such meridional flows. We inject three planets of masses \Mjup{0.5}, \Mjup{1} and \Mjup{2} at 87, 140 and 237~au into a disk model which has been shown to recover the disk thermal structure exceptionally well\cite{Flaherty_ea_2015, Flaherty_ea_2017} (see the methods section). Fig.~\ref{fig:hydro} shows the azimuthally averaged density and kinematic structure after 1.44~Myr, a significant fraction of the age of the system. The derived velocity structure is in excellent agreement with both the observations and previous simulations which only contained a single planet-opened gap\cite{Kley_ea_2001, Szulagyi_ea_2014, Gressel_ea_2013, Morbidelli_ea_2014, Fung_ea_2016, Dong_ea_2019}. This simulation demonstrates that $^{12}$CO emission is able to trace the tops of the meridional flows driven by embedded planets in HD~163296, however the mass of the inferred planets are model dependent and more thorough constraints on these are beyond the scope of this work.

Without the direct detection of the embedded planets opening these gaps, other scenarios are possible. For example, zonal flows are known to drive radial deviations in both $v_{\phi}$ and $v_Z$, allowing also for an efficient cycling of material\cite{Lyra_ea_2008, Johansen_ea_2009, Flock_ea_2015}. In these simulations deviations in $v_{\phi}$ of up to 10\% of the local Keplerian rotation are found and similar values in the vertical direction\cite{Suzuki_Inutsuka_2014}. With the current observations which lack an absolute scaling of the deviations of $v_{\phi}$ or $v_Z$, it is impossible to accurately distinguish between scenarios. The use of different molecular tracers offers a potential solution, allowing for the mapping of the flow structures in regions closer to the disk midplane where they are believed to deviate and thus constrain the origin.

Regardless of the underlying mechanism driving them, the identification of vertical flows with speeds of $\sim$~0.1~$c_s$, suggest a rapid cycling of material from the volatile-rich `molecular layer' of the disk, down towards the considerably cooler midplane which is substantially more shielded from radiation and therefore more chemically inert. It is this volatile-rich gas which will form the atmospheres of the embedded planet, rather than the midplane material which is driven away through the Lindblad torques. Therefore, as the carbon-to-oxygen ratio, C/O, is frequently measured in exoplanetary atmospheres and used a probe of the planet forming location\cite{Oberg_ea_2011, Madhusudhan_2012}, it is the C/O ratio of the in-falling warm molecular regions which should be compared to, rather than the C/O ratio of the disk midplane.

In addition, a radial outflow is found outside $\sim$300~au, marked with the blue dashed box in Fig.~\ref{fig:12COvectors}. Although the absolute scaling of $v_Z$ is dependent on the assumed $v_{\rm LSR}$, the $v_R$ component is measured absolutely so that a true outflow of material is observed, while there is no evidence for such radial motion inwards of 300~au. Viscous spreading of the disk can be ruled out as this should result in much lower velocities on the order of less than \vel{1}, considerably slower than the measured $v_{\rm flow} \sim 30~{\rm m\,s^{-1}}$, however the orientation of the vectors is suggestive that this is the base of a disk wind. A large scale molecular wind, extending over 10\arcsec{} ($\sim$1000~au) and reaching velocities of $\sim 20$km\,s$^{-1}$, has been previously observed in $^{12}$CO around HD~163296 with associated Herbig Haro knots distributed along a jet\cite{Klaassen_ea_2013, Devine_ea_2000}. The large-scale wind kinematics indicate a substantial mass loss rate of 1.4 -- $18.3 \times 10^{-8}~M_{\rm sun}\,{\rm yr}^{-1}$. If this large scale molecular wind is related to the radial outflow we observe, it would suggest disk winds can provide an efficient mechanism to remove angular momentum from the disk without the need for turbulent viscosity, reconciling the inferred low levels of turbulence\cite{Flaherty_ea_2015, Flaherty_ea_2017}. As discussed in the Methods section, the details of the flow structure are dependent on the assumed emission surface and therefore any further analysis of the potential launching mechanisms first requires a more robust measure of the emission surface at the disk edges.

The detection of large-scale flows demonstrates that protoplanetary disks are likely still highly dynamic and actively changing their physical and chemical structures. Looking forward, the ability to infer 3D velocities will aid in further planet detection at (sub-)mm wavelengths and perhaps even aid in understanding the long standing puzzle of disk angular momentum transport.


\clearpage

\clearpage

\clearpage

\renewcommand{\figurename}{Extended Data Fig.}
\setcounter{figure}{0}  

\begin{figure*}
    \centering
    \includegraphics[width=0.6\textwidth]{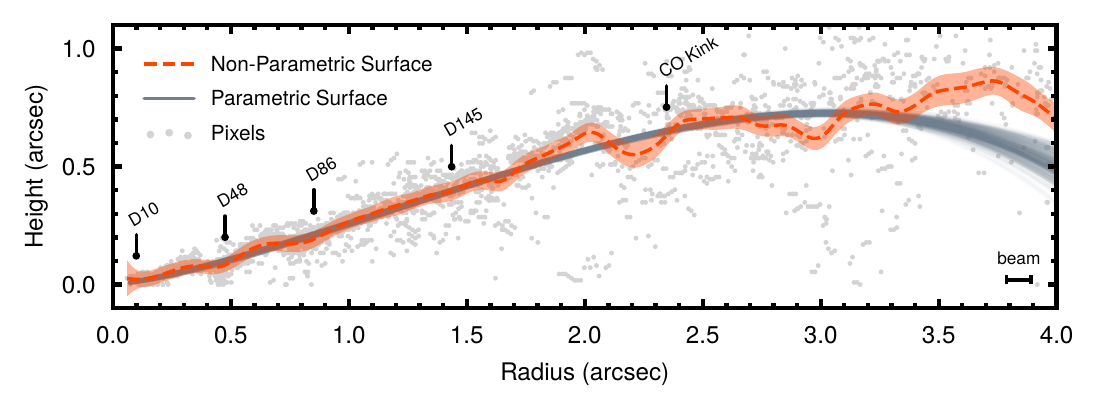}
    \caption{\textbf{A comparison of emission heights inferred for the $^{12}$CO emission.} The gray points in the background represent individual measurements following Pinte et al. (2018a)\cite{Pinte_ea_2018a}, while the red contour shows the Gaussian Process model of this surface including $1\sigma$ uncertainties, as described in Teague et al. (2018a)\cite{Teague_ea_2018a}. Grey lines are random samples from the the parametric fit from modelling the line of sight velocity map, with their spread demonstrating the $1\sigma$ uncertainties. The dust gap locations\cite{Huang_ea_2018b, Isella_ea_2018} and radial location of the velocity perturbation found in $^{12}$CO \cite{Pinte_ea_2018b} are marked. The beam major axis is shown for scale in the bottom right.}
    \label{fig:emission_surface}
\end{figure*}

\begin{figure*}
    \centering
    \includegraphics{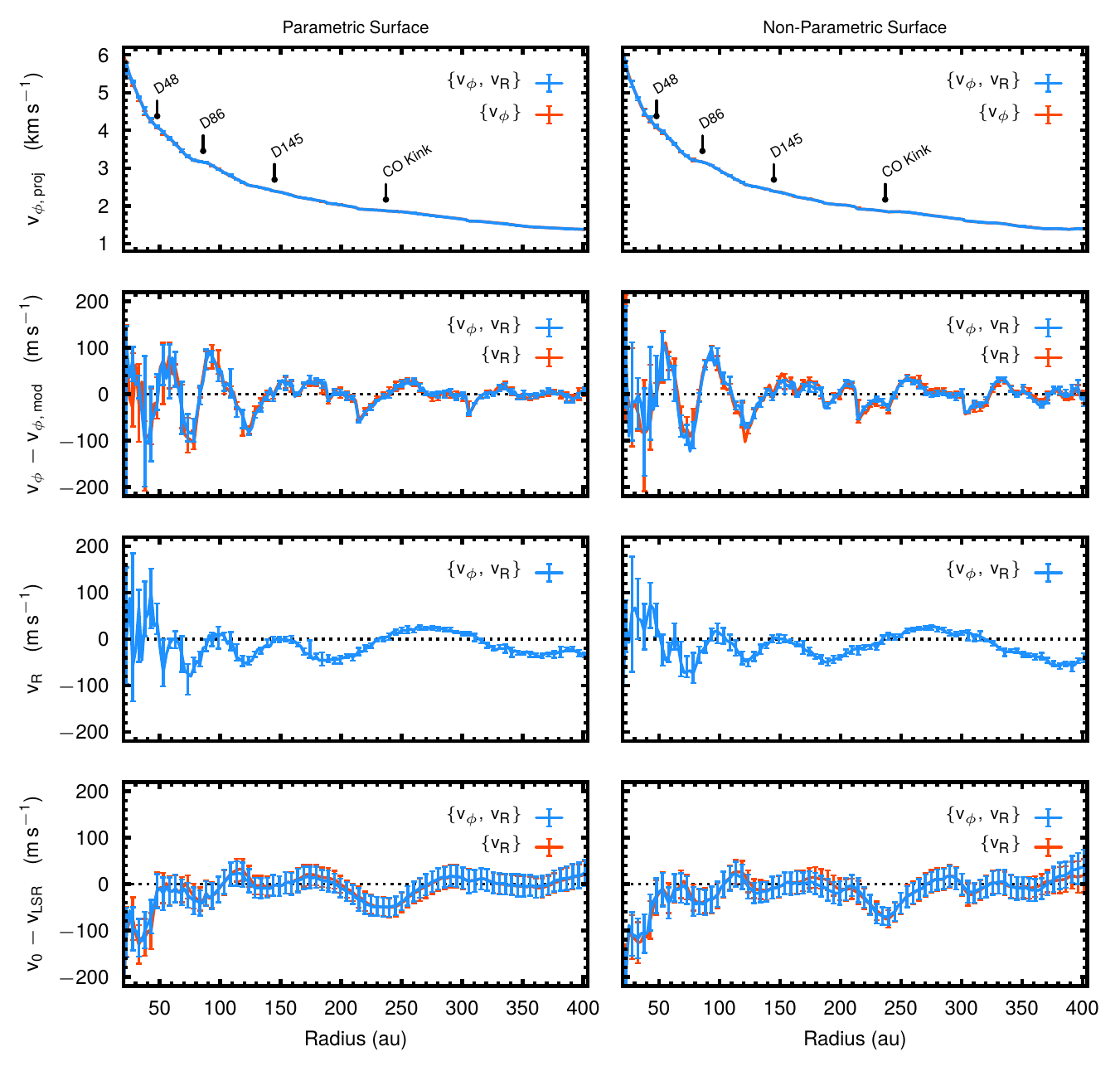}
    \caption{\textbf{Measured velocity structure of HD~163296}. The top panel shows projected rotation velocity $v_{\phi, {\rm proj}}$, while the second panel shows the residual from the 10th-order polynomial fit to $v_{\phi}$ to highlight the small scale structure. The third panel shows the $v_R$ values and the fourth the deviation in the shifted and aligned line centre from the systemic velocity. Velocities in the final three panels have been corrected for projection effects assuming $i = 47.6\degr$. Blue error bars show the inferred velocities assuming both $v_R$ and $v_{\phi}$ components, while values in red assume $v_R = 0~{\rm m\,s}^{-1}$.}
    \label{fig:radialprofiles}
\end{figure*}

\begin{figure*}
    \centering
    \includegraphics[width=\textwidth]{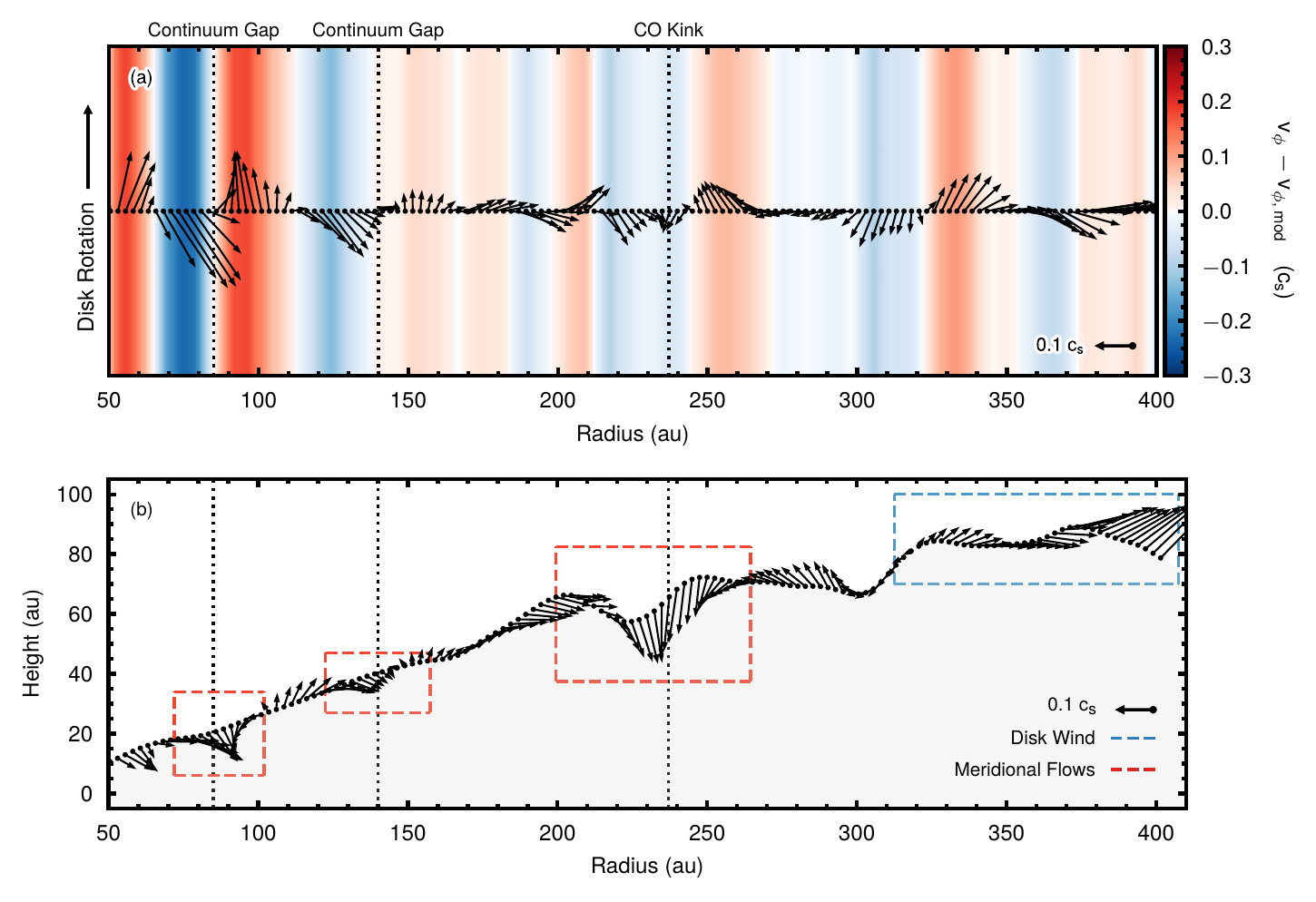}
    \caption{\textbf{Measured velocity structure for the gas in the disk around HD 163296} The same as Fig.~\ref{fig:12COvectors} but using the non-parametric emission surface to deproject the data. Structure in the emission height outside 3\arcsec{} is due to higher noise in the data as described in the text.}
    \label{fig:12COvectors_nonparametric}
\end{figure*}

\begin{figure*}
    \centering
    \includegraphics{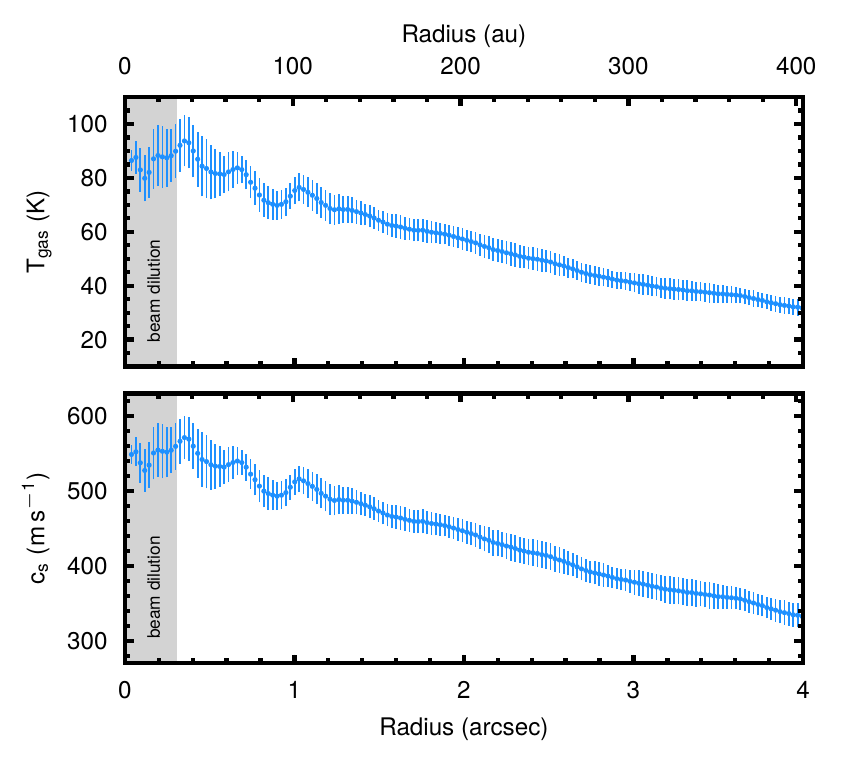}
    \caption{\textbf{Gas temperature and sound speed.} Showing the derived gas temperature, top panel, and the derived gas sound speed, bottom panel. Errorbars show the 1$\sigma$ uncertainty. The drop in these values in the inner $\approx 30$~au is due to beam dilution.}
    \label{fig:gas_temperatures}
\end{figure*}

\begin{figure*}
    \centering
    \includegraphics{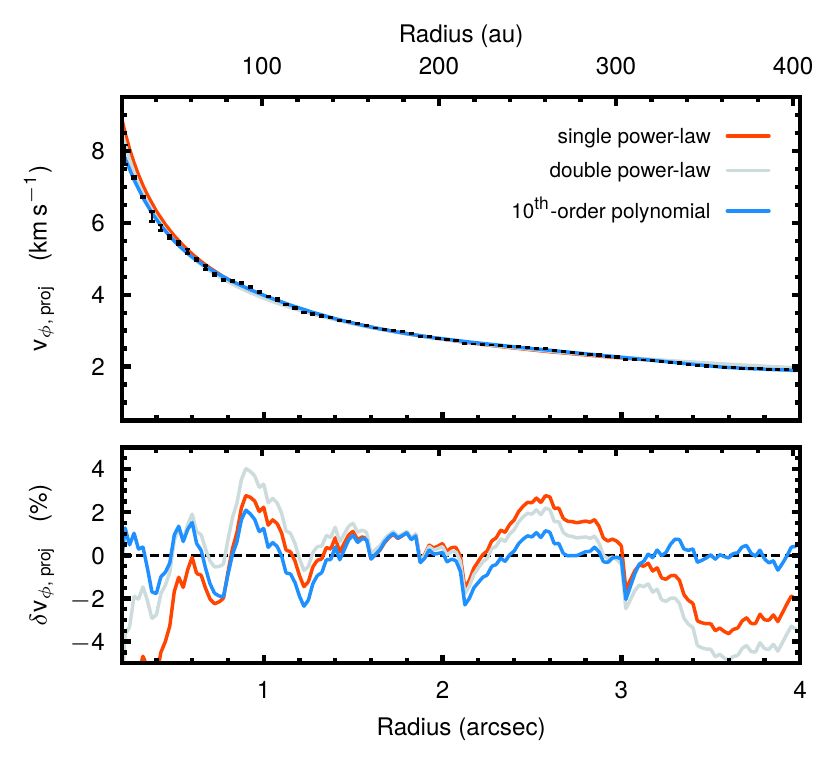}
    \caption{\textbf{Impact of the choice of velocity baseline.} Demonstration of how the choice of $v_{\rm\phi,\, mod}$ affects the residuals from $v_{\phi}$, as in the second row of Fig.~\ref{fig:radialprofiles}. Regardless of the $v_{\rm\phi,\, mod}$ chosen, the structure in $\delta v_{\phi} = (v_{\phi} - v_{\rm\phi,\, mod}) \, / \, v_{\phi}$ persists.}
    \label{fig:vphi_model}
\end{figure*}

\clearpage
\newpage
\begin{methods}

\paragraph{Observations}

We use the publicly available images from the Disk Substructures at High Angular Resolution Program (DSHARP) ALMA large program\cite{Andrews_ea_2018, Isella_ea_2018}. Combining data from three projects, 2013.1.00366.S\cite{Flaherty_ea_2015}, 2013.1.00601.S\cite{Isella_ea_2016} and 2016.1.00484.L\cite{Andrews_ea_2018}, a spatial resolution of $0.104\arcsec \times 0.095\arcsec$ (corresponding to $10.5~{\rm au} \times 9.6~{\rm au}$ at the source distance of 101~pc) was achieved and a spectral resolution of \vel{640} after Hanning smoothing. More details about the data reduction and imaging procedure can be found in Isella et al. (2018).

\paragraph{Deriving an Emission Surface}

In order to accurately deproject the observations into annuli of constant radius we must take into account the flared emission surface of the disk. We use two different approaches in order to derive a robust emission surface. The first approach uses the morphology of the map of the line center. For a geometrically thin disk, the rotation map will have a symmetric dipole pattern. However, with an elevated emission surface, the red- and blue-shifted lobes of the dipole bend away from the disk semi-major axis, allowing for a constraint on the emission surface. We first make a map of the line centres using \texttt{bettermoments}\cite{Teague_Foreman-Mackey_2018} which fits a quadratic curve to the pixel of peak intensity and the two neighbouring pixels. This has the significant advantage over the more traditionally used intensity-weighted averaged velocities (first moment maps) in that no clipping is required and that the near side of the disk is readily distinguished from the far side, in addition to returning well characterised uncertainties on the line center. The resulting map of the line centers is shown in Fig.~\ref{fig:12COv0}a.

Assuming that the kinematics of the disk are dominated by Keplerian rotation, we also fit the rotation map with a Keplerian rotation pattern including the correction for the height above the midplane\cite{Rosenfeld_ea_2013},

\begin{equation}
    v_0 = \sqrt{\frac{GM_{\rm star}R^2}{(R^2 + Z^2)^{3/2}}} \cdot \cos \theta \cdot \sin (i)
\end{equation}

\noindent where $R$ and $Z$ are the midplane radius and the height above the midplane, respectively, $\theta$ is the polar angle of the disk and $i$ is the inclination of the disk. The emission surface is parameterised as two power-law profiles to capture the flared increasing surface over most of the disk, but also the drop in the outer disk due to the decreasing gas surface density,

\begin{equation}
    Z(R) = Z_0 \times (R \, / \, 1\arcsec)^{\varphi} - Z_1 \times (R \, / \, 1\arcsec)^{\psi}.
\label{eqn:emission_surface}
\end{equation}

\noindent where both $Z_0$ and $Z_1$ should be positive. In addition to the emission surface, we let the source center and disk inclination and position angle, $\{x_0,\, y_0,\, i,\, {\rm PA}\}$, vary as well as a variable stellar mass, $M_{\rm sun}$ and systemic velocity, $v_{\rm LSR}$, leaving 10 free parameters for the model.

The calculation of the posterior distributions was performed using the Python package \texttt{eddy}\cite{eddy} assuming a source distance of 101~pc\cite{Bailer-Jones_ea_2018}. Only spatially independent pixels, those at least a beam FWHM apart, were considered in the calculation of the likelihood.  Flat priors were assumed for all values except the inclination which has a prior of $i = 46.7\degr \pm 0.1\degr$ based on the fits to the continuum observations\cite{Isella_ea_2018}. 64 walkers were run for 10,000 steps for an initial burn-in period, before an additional 10,000 steps were used to sample the posterior distributions. The resulting posterior distributions were found to be $x_0 = -5_{-2}^{+2}~{\rm mas}$, $y_0 = 3_{-2}^{+2}~{\rm mas}$, $i = 46.8_{-0.1}^{+0.1}$, ${\rm PA} = 312.83_{-0.02}^{+0.02}\degr$, $M_{\rm star} = 2.022_{-0.004}^{+0.004}~M_{\rm sun}$, $v_{\rm LSR} = 5.763_{-0.001}^{+0.001}~{\rm km\,s^{-1}}$ and an emission surface described by $Z_0 = 265_{-3}^{+3}~{\rm mas}$, $\varphi = 1.29_{-0.03}^{+0.04}$, $Z_1 = 6_{-3}^{+2}~{\rm mas}$ and $\psi = 3.8_{-0.3}^{+0.3}$, where the uncertainties are the 16th to 84th percentiles of the posterior distribution about the median value. We note these uncertainties represent only the statistical uncertainty on the values and the systematic uncertainties, for example due to neglecting the pressure correction term in the calculation of the velocity profile, are not included. Fixing $Z_1 = 0$ results in a comparable geometric parameters and emission surface characterized by $Z_0 = 280_{-2}^{+2}~{\rm mas}$, but a smaller $\varphi = 0.994_{-0.007}^{+0.007}$ to account for the drop in the emission surface at larger radii, broadly consistent with previous work\cite{Isella_ea_2018}.

The second method follows the method presented in Pinte et al. (2018a)\cite{Pinte_ea_2018a} which allows for a non-parametric surface which can account for a drop in the emission surface into gap regions, as previously demonstrated for C$^{18}$O emission in this source\cite{Teague_ea_2018a}. This method assumes that the gas is on circular orbits and uses the asymmetry of the emission relative to the major axis of the disk to infer the height of the emission relative to the disk midplane. For this approach we adopt the inclination and position angle found from continuum\cite{Isella_ea_2018}, $i = 46.7\degr$ and ${\rm PA} = 133.3\degr$ (note here there is a 180\degr{} degeneracy in the PA when modelling continuum emission). Following Teague et al. (2018a)\cite{Teague_ea_2018a}, the raw data points were modelled with a Gaussian Process resulting in a smooth emission surface, rather than radially binning the data.

Extended data Figure~\ref{fig:emission_surface} compares the two derived emission surfaces. The grey solid lines show 50 random samples of the parametric surface, while the red dashed line shows the Gaussian Process model and the associated 3$\sigma$ uncertainties when modelling the raw points shown as grey dots. In the outer disk, $r \gtrsim 3\arcsec$, the two surfaces deviate. This is because for the non-parametric method, the fit is dependent on individual pixels which have a lower signal-to-noise ratio in the outer disk (as demonstrated by the significant scatter in the grey points). Conversely, as the parametric model is less flexible, it is therefore less affected by the noisy outer disk. Using only the data points for $r < 3\arcsec$ we find a consistent non-parametric emission surface.

\paragraph{Measuring the Velocity Structure}

To infer radial profiles of the radial and rotational velocities, $v_R$ and $v_{\phi}$, respectively, we follow the method outlined in Teague et al. (2018a,b)\cite{Teague_ea_2018a, Teague_ea_2018b}, using the Python implementation \texttt{eddy}\cite{eddy}. The disk is first split into annuli of constant radius accounting for the inferred emission surface, each 25~mas wide (roughly 1/4 of the beam FWHM) from which spatially uncorrelated pixels (at least a beam FWHM apart) were randomly selected (the pixel scaling was 13~mas). The line center of each pixel was assumed to be the sum of the projected component of $v_{\phi}$, and $v_R$ such that

\begin{equation}
    v_{\rm LOS}(\phi) = v_{\phi,\, {\rm proj}} \cdot \cos \phi + v_{R,\, {\rm proj}} \cdot \sin \phi
\end{equation}

\noindent where $\phi$ is the deprojected polar angle along the annulus. The best fitting $v_{\phi}$ and $v_R$ values were those which maximised the likelihood when a Gaussian Process was fit to the shifted and stacked data which is a robust measure of the alignment of the spectra, as described in Teague et al. (2018b)\cite{Teague_ea_2018b}. For each annulus, 20 different realisations with different pixels were performed and the posterior distributions combined.

To calculate the projected vertical component of the velocity, $v_{Z,\, {\rm proj}}$, we fit a Gaussian to the aligned and stacked spectra using the best-fit $v_{\phi,\, {\rm proj}}$ and $v_{R,\, {\rm proj}}$ values. To improve the quality of the Gaussian fit, we resample the stacked spectra to \vel{40}, a factor of 16 higher than the original data. This resampling will not alter the intrinsic line profile and will preserve any systematic effects, such as the spectral response function of the ALMA correlator and the Hanning smoothing. While these systemic effects will bias measurements of the line peak or width, the line center should remain unaffected. We perform two fits to the data. First using the whole spectrum to identify the line center and a rough line width, then a second only the line core defined as where $|v - v_0| \leq \Delta V$, in order to remove biases due to the noisy line wings arising due to contamination from the far side of the disk (see the Appendix in Teague et al. (2018b)\cite{Teague_ea_2018b}.) The measured line center is then considered the sum of the systemic velocity, $v_{\rm LSR}$ and the projected vertical velocity, $v_{Z,\, {\rm proj}}$. Without a precise measure of $v_{\rm LSR}$, the inferred $v_{Z}$ radial profiles are therefore only relative values. In future, observations of an optically thin line where $v_Z$ components would cancel out from either side of the disk may allow for a better determination of the $v_{\rm LSR}$ and constraints on the absolute values of $v_{z,\, {\rm proj}}$. Finally, the projected velocity components were corrected for their projection using a disk inclination of $46.7\degr$.

We perform this method for both of the emission surfaces we have derived to verify that the deprojection does not significantly change the derived velocity vectors. Furthermore, to check that the inclusion of a $v_R$ term was not changing the inferred $v_{\phi}$ value, we repeat the procedure fixing $v_{R, {\rm proj}} = 0$ and find no change in the inferred $v_{\phi,\, {\rm proj}}$. This is expected because the radial and rotational velocity components are orthogonal and thus $v_R$ components cannot correct for changes in $v_{\phi}$ and vice versa. We also use the differences between the two methods to estimate the systematic uncertainties on the derived velocities. Across the whole radius of the disk, the average differences between the the methods are $1.5\sigma$, $1\sigma$ and $0.2\sigma$ for $v_{\phi}$, $v_R$ and $v_Z$, respectively. Within the gap regions where the two surfaces are most different, most notably D48 or at 220~au, these rise to $\sim 3\sigma$ for all three components. Despite the difference in the emission heights, the qualitative flow structures remain, although the wind signature is less well defined than with the smooth parametric surface.

As the perturbations are expected to be on the order of a few percent of the Keplerian rotation, we must subtract a background profile from $v_{\phi}$ in order to clearly see them. However, as $v_{\phi}$ is a combination of the Keplerian rotation which varies depending on the height of the emission surface, in addition to perturbations from the radial pressure gradient and self-gravity of the disk\cite{Rosenfeld_ea_2013}, there is no simple predictive fit to the curve. To act as a model of the background rotation, $v_{\phi,\, {\rm mod}}$, we fit a 10th-order polynomial to $v_{\phi}$. We have also tried single and broken power-law functions, but these were unable to reliably capture the overall trend at small and large radii. Extended data Figure~\ref{fig:vphi_model} demonstrate the choice of $v_{\phi,\, {\rm mod}}$ and how the resulting residuals are affected. We stress that this is for presentation purpose only and, if any quantitative fits were to be made, they should be made directly to $v_{\phi}$, as in previous works\cite{Teague_ea_2018a, Teague_ea_2018b}.

The resulting velocities are plotted in Extended data Figure~\ref{fig:radialprofiles} in blue, with the three components in panels (b) -- (d) corrected for the projection. The values in red show the results when we fix $v_R = 0$. The locations of the gaps in the continuum\cite{Huang_ea_2018b, Isella_ea_2018} and previously reported planets\cite{Teague_ea_2018a, Pinte_ea_2018b} are annotated in the top row. The results for both types of emission surface are consistent, demonstrating that the results are robust to the choice of emission surface.

To convert the velocities to Mach numbers, we use the peak value from the aligned and deprojected spectra to measure the local gas temperature using the full Planck law and calculate a local sound speed. The derived gas temperatures, consistent with previous analyses\cite{Isella_ea_2018}, and derived sound speeds are shown in Extended data Figure~\ref{fig:gas_temperatures}. This temperature may be underestimated if the emission line is only marginally thick, $\tau \lesssim 5$, however at this temperature this requires a column density of $N({\rm CO}) \lesssim 10^{16}~{\rm cm^{-2}}$, considerably less than found in previous models of CO isopologue emission in this source\cite{Flaherty_ea_2015, Flaherty_ea_2017}. Due to the low spectral resolution of the data and confusion from the far side of the disk, measuring the line width will not provide a reasonable measure of the gas kinetic temperature\cite{Teague_ea_2016, Teague_ea_2018b}.

\paragraph{Hydrodynamic Simulation Setup}

We solve the isothermal hydrodynamic continuity equations and the Navier-Stokes equations in three-dimensional spherical $(r, \theta, \phi)$ coordinates using publicly available planet-disk interaction code FARGO3D\cite{benitez16}, adopting the orbital advection algorithm\cite{masset00}.

We adopt a parametric disk model similar to the one presented in Flaherty et al. (2017)\cite{Flaherty_ea_2017} which is shown to well reproduce CO isotopologue emission. The vertically integrated surface density of the disk follows

\begin{equation}
\label{eqn:density}
    \Sigma_{\rm gas} (R) = \frac{M_{\rm gas}}{2 \pi R_c^2} (2-\gamma) \left( \frac{R}{R_c} \right)^{-\gamma} \exp{\left[ -\left( \frac{R}{R_c}\right)^{2-\gamma} \right]},
\end{equation}

\noindent where $M_{\rm gas} = 0.09~M_\odot$ is the total gas mass, $R_c=160.6$~AU is a characteristic radius, and $\gamma=1$. The disk temperature structure has a power-law profile with cylindrical radius $R$, and a vertical gradient is imposed at each radius to smoothly connect the cold midplane and warm surface as follows.

\begin{eqnarray}
\label{eqn:temperature}
T(R, Z) =
    \begin{cases}
        T_{\rm atm}(R) + (T_{\rm mid}(R) - T_{\rm atm}(R)) \cos^2\left( \frac{\pi}{2} \frac{Z}{Z_q} \right)& \text{if}~Z < Z_q \\
        T_{\rm atm}(R) & \text{if}~Z \geq Z_q \\
    \end{cases}
\label{eqn:smix}
\end{eqnarray}

\noindent Here, $T_{\rm mid}(R) = 16.6~{\rm K}~(R/R_c)^q$ and $T_{\rm atm}(R) = 88.9~{\rm K}~(R/R_c)^q$ where $q=-0.216$, and $Z_q = 97.8 (R/R_c)^{1.3}$~au.

Using the parametric disk model describe above, we solve the vertical hydrostatic equilibrium equation to obtain an initial density distribution:

\begin{equation}
    \rho(R,Z) = \rho(R,0) \frac{c^2(R,0)}{c_s^2(R,Z)} \exp \left[ - \int_{0}^{Z} {\frac{1}{c_s^2(R,Z')}  \frac{GM_*Z'}{(R^2 + Z'^2)^{3/2}}} {\rm d}Z' \right],
\end{equation}

\noindent where $\rho$ is the gas density, $c_s$ is the sound speed, $G$ is the gravitational constant, and $M_*$ is the stellar mass. We numerically find the density at each grid cell such that the vertically integrated surface density at the radius becomes the one given in Equation (\ref{eqn:density}). With the temperature and density assigned for each grid cell, we compute the initial angular velocity that satisfies the radial force balance taking into account the gas pressure gradient:

\begin{equation}
    \Omega = \left[ \Omega_K^2 \sin\theta + \frac{1}{\rho r \sin^2\theta} \frac{\partial P}{\partial r} \right ]^{1/2}.
\end{equation}

\noindent In the above equation $\Omega_K = \sqrt{GM_*/R^3}$ is the Keplerian angular velocity and $P=\rho c_s^2$ is the gas pressure. The initial radial and meridional velocities are set to zero.

The simulation domain extends from $r_{\rm in} = 32.1$~au to $r_{\rm out} = 401.5$~au in radius and from 0 to $2\pi$ in azimuth. In the meridional direction, we include only the upper half of the disk assuming a symmetry across the disk midplane. The meridional domain extends from the midplane ($\theta=\pi/2$) to 20.6 degrees above the midplane ($\theta=\pi/2-0.36$). We adopt 256 logarithmically-spaced grid cells in the radial direction, 36 uniformly-spaced grid cells in the meridional direction, and 635 uniformly-spaced grid cells in the azimuthal direction.

We add 0.5 Jupiter-mass planet at 87~au, 1 Jupiter-mass planet at 140~au, and 2 Jupiter-mass planet 237~au. The planet masses and radial locations are chosen based on previous modelling that successfully reproduced ALMA continuum and CO line observations \cite{Liu_ea_2018,Teague_ea_2018a,Pinte_ea_2018b}. The simulation runs for 1.44~Myr, covering a significant fraction of the system's age. This corresponds to 1000 orbits at $R=R_c$. We linearly increase planet masses during the first 0.144~Myr. We implement a uniform disk viscosity of $\alpha=10^{-3}$, consistent with the constraints made with molecular line observations of the disk \cite{Flaherty_ea_2015,Flaherty_ea_2017}.

\paragraph{Additional Hydrodynamic Simulations}

We carry out the fiducial calculation at a factor of two higher numerical resolution (i.e., $512\times72\times1270$ grid cells) for a shorter duration of 0.432~Myr. We confirm that the meridional circulation pattern and its strength are not affected by the change in numerical resolution.

While previously inferred disk thermal profile and planet masses successfully reproduce the overall circulatory flow patterns, our fiducial calculation exhibits stronger rotational motions than vertical motion around the gaps compared with those inferred from the observation. We carry out addition calculations with (1) a factor of two smaller planet masses and (2) a factor of two smaller disk surface temperature ($T_{\rm atm}$ in Equation \ref{eqn:temperature}), but these changes result in lowering the level of both rotational and vertical motions. One possible way to have smaller rotational velocity deviations around gaps, while maintaining vertical motions, is if the Reynolds and/or Maxwell stresses in the surface layers are larger than those near the midplane. In such a case, restoring flows near the surface would fill in a gap more efficiently, reducing the pressure gradient across the gap and hence rotation velocity deviations.

The vertical motion from the top of a gap to the planet is, on the other hand, less sensitive to the viscous stress\cite{Morbidelli_ea_2014}; instead, thermal/magnetic properties within the gap can have larger influence on the flow pattern/speed\cite{Gressel_ea_2013}. Including ideal/non-ideal magneto-hydrodynamic effects and more realistic thermal structure via Monte Carlo radiative transfer calculations would be required to examine such a possibility. However, these simulations would require sufficiently high numerical resolution to capture unstable modes of the magneto-rotational instability\cite{Balbus_ea_1998} and long enough integration time for planets to open gaps in a global simulation domain, and are computationally prohibitive for the moment.

\begin{addendum}
    \item [Data availability] This paper makes use of the following ALMA data: ADS/JAO.ALMA\#2013.1.00366.S, ADS/JAO.ALMA\#2013.1.00601.S, and ADS/JAO.ALMA\#2016.1.00484.L. The raw data is available from the ALMA archive (\url{http://almascience.nrao.edu/aq/}), while the imaged data and scripts are available from the DSHARP website (\url{https://bulk.cv.nrao.edu/almadata/lp/DSHARP/}). The Python packages used for the analysis of the data are available via their GitHub repositories: \texttt{bettermoments} (\url{https://github.com/richteague/bettermoments}) and \texttt{eddy} (\url{https://github.com/richteague/eddy}).

    \item [Acknowledgements] This paper makes use of the following ALMA data: ADS/JAO.ALMA\#2013.1.00366.S, ADS/JAO.ALMA\#2013.1.00601.S, and ADS/JAO.ALMA\#2016.1.00484.L. ALMA is a partnership of European Southern Observatory (ESO) (representing its member states), National Science Foundation (USA), and National Institutes of Natural Sciences (Japan), together with National Research Council (Canada), National Science Council and Academia Sinica Institute of Astronomy and Astrophysics (Taiwan), and Korea Astronomy and Space Science Institute (Korea), in cooperation with Chile. The Joint ALMA Observatory is operated by ESO, Associated Universities, Inc/National Radio Astronomy Observatory (NRAO), and National Astronomical Observatory of Japan. The National Radio Astronomy Observatory is a facility of the National Science Foundation operated under cooperative agreement by Associated Universities, Inc. R.T and E.A.B acknowledge funding from NSF grant AST-1514670 and NASA grant NNX16AB48G. J.B. acknowledges support from NASA grant NNX17AE31G, and computing resources provided by the NASA High-End Computing (HEC) Program through the NASA Advanced Supercomputing (NAS) Division at Ames Research Center and by the Extreme Science and Engineering Discovery Environment (XSEDE) which is supported by National Science Foundation grant number ACI-1548562.

    \item[Author Contributions] R.T. devised the method and analysed the data. J.B. ran the hydrodynamic simulations. All authors wrote the manuscript and were participants in the discussion and interpretation of results, determination of the conclusions and revision of the manuscript.
    
    \item[Author Information] Reprints and permissions information is available on request. The authors declare no competing financial interests. Readers are welcome to comment on the online version of the paper. Correspondence and requests for materials should be addressed to R.T.~(email: rteague@umich.edu).
\end{addendum}

\end{methods}


\clearpage


\end{document}